\documentstyle[aps,preprint,tighten,epsf]{revtex} 
\begin{document}
\preprint{imsc/98/06/29}
\draft
\title{Dual gluons and monopoles in 2+1 dimensional Yang-Mills theory} 
\author{Ramesh Anishetty \thanks{e-mail:ramesha@imsc.ernet.in}\and 
Pushan Majumdar \thanks{e-mail:pushan@imsc.ernet.in}
\and H.S.Sharatchandra \thanks{e-mail:sharat@imsc.ernet.in}} 
\address{Institute of Mathematical Sciences,C.I.T campus Taramani.  
Madras 600-113} 
\maketitle
\begin{abstract} 
2+1-dimensional Yang-Mills theory is reinterpreted in
terms of metrics on 3-manifolds. The dual gluons are related to 
diffeomorphisms of the
3-manifolds. Monopoles are identified with points where the Ricci
tensor has triply degenerate eigenvalues. The dual gluons have the
desired interaction with these monopoles. This would give a mass for the
dual gluons resulting in confinement.  
\end{abstract} 
\section{Introduction}

Quark confinement is well understood in 2+1 dimensional compact U(1) gauge 
theory. 
It is a consequence of the existence of a monopole plasma \cite{P1}\cite{P2}. 
Duality transformation \cite{BKM} turned out to be very useful in this
context. It is of interest to know how far these ideas can be extended 
to non-abelian gauge theories. For this 
reason, duality transformation for 2+1-dimensional Yang-Mills theory was 
obtained in lattice 
gauge theory in both hamiltonian \cite{AH} and partition function \cite{M} 
formulations. The dual theory exhibits close relationship to
2+1-dimensional gravity, but without diffeomorphism invariance. This also 
indicates a way of describing 
the dynamics using local gauge invariant variables.

In this paper, we consider duality transformation for 2+1-dimensional 
(continuum) Yang-Mills theory in close analogy to the case of compact U(1) 
lattice gauge theory \cite{BKM}. We reinterpret the Yang-Mills theory as a 
theory of 3-manifolds, as in gravity, but without diffeomorphism
invariance.
We use this relation for identifying the dual gluons and 
their interactions. The dual gluons are related to diffeomorphisms
of the 3-manifold. We also identify the monopoles in the dual theory. 't Hooft 
\cite{HOO} has advocated the use of a composite 
Higgs to locate the monopoles. Here we propose to use the orthogonal set of 
eigenfunctions of a gauge invariant, (symmetric) local, matrix-valued field 
for this purpose. Isolated points where the eigenvalues 
are triply degenerate have topological significance and they locate the 
monopoles. We use the Ricci tensor to construct a new coordinate 
system for the 3-manifold. The monopoles 
are located at the singular points of this coordinate system and they have 
the expected interactions with the dual gluons. We expect that these 
interactions lead to a mass for the dual gluons and result in 
confinement as in the U(1) case. 

Lunev \cite{Lun} has pointed out the relationship of 2+1-dimensional 
Yang-Mills theory 
with gravity. He uses a gauge invariant composite $B_{i}^{a}B_{j}^{a}$ as a 
metric, and rewrites the classical Yang-Mills dynamics for it. The 
corresponding formulation of the quantum theory is somewhat involved. Our 
metric is in a sense dual of Lunev's choice. As we make formal transformations 
in the functional integral, the quantum theory is simpler and has a nicer 
interpretation.

There are also approaches to relate 3+1-dimensional Yang-Mills theory to a 
theory of a metric \cite{Haag}. On the other hand, the dual theory in 
3+1-dimensions can also be related to a new SO(3) gauge theory \cite{MS1}.

In section 2 we briefly review duality transformation and confinement in 
2+1-dimensional compact U(1) lattice gauge theory. In section 3 we obtain the 
dual description of 2+1-dimensional Yang-Mills theory in close analogy to 
section 2. We point out the close relationship to gravity and identify the
dual 
gluons and their interactions. In section 4 we provide a new characterization 
of monopoles using eigenfunctions of the symmetric matrix 
$B_{i}^{a}B_{i}^{b}$. In section 5 we use the Ricci tensor to construct a 
preferred coordinate system for 3-manifolds. We relate the monopoles to 
singularities of this coordinate system. We also identify their 
interactions with the dual gluons. Section 6 contains our conclusions.

\section{Review of confinement in 2+1-dimensional compact U(1) Lattice Gauge 
Theory}

In this section we briefly review duality transformation \cite{BKM} and 
confinement \cite{P1} in 
2+1 dimensional compact U(1) lattice gauge theory. This provides a
paradigm for our analysis of 2+1 dimensional Yang-Mills case.

The Euclidean partition function in the Villain formulation is given by 
\begin{equation}\label{part}
Z=\sum_{h_{ij}}\prod_{ni}\int_{-\infty}^{\infty}\!dA_{i}(n)
\;exp \:(-\frac{1}{4\kappa^{2}}{\sum_{nij}\:
[\bigtriangleup_{i}A_{j}(n) 
-\bigtriangleup_{j}A_{i}(n)+h_{ij}(n)]^{2}}).
\end{equation}
Here $A_{i}(n)\in (-\infty,\infty)$ are non-compact link variables on 
links joining the sites $n$ and $n+\hat{i}$. $h_{ij}(n)=0,\pm 1, \pm 2 
\ldots$ are integer variables corresponding to the monopole degrees of 
freedom and are associated with the plaquette $(n\hat{i}\hat{j})$.
$\bigtriangleup_{i}$ is the difference operator, $\bigtriangleup_{i}\phi(n)
=\phi(n+\hat{i})-\phi(n)$. 
We may introduce an auxiliary variable $e_{i}(n)$ to rewrite $Z$ as
\begin{equation}\label{auxpart}
Z=\sum_{h_{ij}}\prod_{ni}\int_{-\infty}^{\infty}\!dA_{i}(n)\int_{-\infty}
^{\infty}de_{i}(n)\;exp\left(-\sum_{ni}\:[e_{i}(n)]^{2}
+\frac{2i}{\kappa}\sum_{nij}\epsilon_{ijk}\:e_{k}(n)[\bigtriangleup_{i}A_{j}(n) 
+\frac{1}{2}h_{ij}(n)]\right).
\end{equation}
Integration over $A_{j}(n)$ gives the $\delta$ function constraint 
\begin{equation}\label{constr}
\epsilon_{ijk}\bigtriangleup_{j}e_{k}(n)=0
\end{equation}
for each $n$ and $\hat{i}$. The solution is 
$e_{i}(n)=\bigtriangleup_{i}\phi(n)$.
Thus we get the dual form of the partition function
\begin{equation}\label{dualpart}
Z=\sum_{h_{ij}}\prod_{ni}\int_{-\infty}^{\infty}\!d\phi(n)\;exp\sum_{n}\left(
-[\bigtriangleup_{i}\phi(n)]^{2}+\frac{i}{2\kappa}\phi(n)\rho(n)\right),
\end{equation}
where $\rho(n)=\frac{1}{2}\epsilon_{ijk}\bigtriangleup_{i}h_{jk}(n)$.
This has the following interpretation. The field $\phi$ describes the dual 
photon. (In 2+1 
dimensions, the photon has only one transverse degree of freedom and 
this is captured by the scalar field $\phi(n)$). The monopole 
number at site $n$ is given by $\rho(n)$. 
It takes integer values and the dual photon couples locally 
to it with strength $1/\kappa$.

If we sum over the monopole degrees of freedom, we get a mass term for 
$\phi(n)$ \cite{P1}\cite{BKM}. The reason for this is that the monopole 
plasma is screening the long range interactions 
between the monopoles. A Wilson loop for the electric charges in this 
system would correspond to a dipole sheet in this plasma. This gives an 
area law and hence a linear confining potential between static electric 
charges.  

The advantage of this formal duality transformation is that it gives a 
precise separation of the `spin wave' and the `topological' degrees of 
freedom. Therefore it provides a stepping stone for going beyond 
semi-classical approximations.

We use this approach for 2+1 dimensional Yang-Mills theory in the next
section.
\section{Dual gluons in 2+1-dimensional Yang-Mills theory}

In this section we point out the close relationship between Yang-Mills 
theory and Einstein-Cartan formulation of gravity in 2+1 (or 3 Euclidean) 
dimensional space. We use this analogy extensively throughout the paper.

The Euclidean partition function of 2+1 dimensional Yang-Mills theory is 
\begin{equation}\label{contpart}
Z=\int {\cal D}A_{i}^{a}(x)\:exp\left(-\frac{1}{2\kappa^{2}}\int 
d^{3}x B_{i}^{a}(x) B_{i}^{a}(x)\right)
\end{equation}
where $\{A_{i}^{a}(x),\;\;(i,a\:=1,2,3)\}$ is the Yang-Mills potential and 
\begin{equation}\label{strength} 
B_{i}^{a}=\frac{1}{2}\epsilon_{ijk}(\partial_{j}A_{k}^{a}-
\partial_{k}A_{j}^{a}+\epsilon^{abc} A_{j}^{b} A_{k}^{c})
\end{equation}
is the field strength.
As in section 2, we rewrite Z as \cite{M}
\begin{equation}\label{auxcontpart}
Z=\int {\cal D}A_{i}^{a}(x)\;{\cal D}e_{i}^{a}(x)\;
exp\left\{\int d^{3}x (-\frac{1}{2}[e_{i}^{a}(x)]^{2}+\frac{i}{\kappa} 
e_{i}^{a}(x)B_{i}^{a}(x))\right\}.
\end{equation}
The second term in the exponent is precisely the Einstein-Cartan action 
for gravity in 3 (Euclidean) dimensions. $e_{i}^{a}(x)$ is the driebein 
and $\omega_{i}^{ab}=\epsilon^{abc}A_{i}^{c}$ the connection 1-form.

In contrast to section 2, we do not get a $\delta$ function constraint 
on integrating over $A_{i}^{a}$ in this case. 
Since $A$ appears at most quadratically in the exponent, the integration over 
$A$ may be explicitly performed. This integration is equivalent to solving 
the classical equations of motion for $A$ as a functional of $e$ and
replacing $A$ by this solution :
\begin{equation}\label{torsionfree}
\epsilon_{ijk}(\partial_{j}\delta^{ac}+\epsilon_{abc}A_{j}^{b}[e])e_{k}^{c}
(x)=0.
\end{equation}
Now (\ref{torsionfree}) is 
precisely the condition for a driebein $e$ to be torsion free with respect to 
the connection 1-form $A_{i}^{c}$. 

If we assume the $3\times 3$ matrix $e_{i}^{a}$ to be non-singular, 
then this solution $A[e]$ can be explicitly given \cite{MS2}.
In this case, no information is lost by multiplying (\ref{torsionfree}) by 
$e_{l}^{a}$ and summing over $a$. We get,
$\epsilon_{ijk}e_{l}^{a}\partial_{j}e_{k}^{a}+ 
|e|(e^{-1})^{m}_{b}\epsilon_{klm}\epsilon_{ijk}A_{j}^{b}[e]=0.$
Defining $A_{j}^{b}(e^{-1})_{b}^{m}=A_{jm}$, we get,
$A_{li}[e]-\delta_{li}A_{mm}[e]=(1/|e|)\epsilon_{ijk}e_{l}^{a}\partial_{j}
e_{k}^{a}.$
Taking the trace on both sides,
$A_{mm}[e]=-(1/2|e|) \epsilon_{ijk}e_{i}^{a}\partial_{j}e_{k}^{a}.$
Then, $A_{l}^{b}[e]=\frac{e_{i}^{b}}{|e|}\left( 
\epsilon_{ijk}e_{l}^{a}\partial_{j}e_{k}^{a}
-\frac{1}{2}\delta_{li}\epsilon_{mjk}e_{m}^{a}\partial_{j}e_{k}^{a}\right).$

By a shift of $A$, $A=A[e] + A^{\prime}$, the integration over $A$ reduces to
\begin{equation}\label{shift}
\int{\cal
D}A^{\prime}\:exp\left(\frac{i}{\kappa}\int\;A_{ia}^{\prime}
e_{ia,jb}A_{jb}^{\prime}\right)=\frac{1}{det^{1/2}(e_{ia,jb})}
=\frac{1}{det^{3/2}(e_{i}^{a})},
\end{equation}
where $e_{ia,jb}=\epsilon_{ijk}\epsilon^{abc}e_{k}^{c}$.

$B_{i}^{a}$ is related to the Ricci tensor $R_{ik}$ as follows:
\begin{equation}\label{ric}
R_{ik}=F_{ij}^{ab}e_{k}^{a}(e^{-1})^{j}_{b}
\end{equation}
where $F_{ij}^{ab}=\epsilon_{ijk}\epsilon^{abc}B_{k}^{c}$. Thus an integration 
over $A$ gives,
\begin{equation}\label{grav}
Z=\int\:{\cal D}g\:exp\left( -\frac{1}{2}g_{ii}+\frac{i}{\kappa}\sqrt{g}R\right)
\end{equation}
where the metric $g_{ij}=e_{i}^{a}e_{j}^{a}$ and $R=R_{ik}g^{ki}$. Note that 
${\cal D}g={\cal D}e\:det^{-3/2}(e_{i}^{a})$, as required.  The 
configurations where $e$ is singular is naively a set of measure zero, 
so that the assumption $|e|\neq 0$ is reasonable.

Equation (\ref{grav}) provides a reformulation of 2+1-dimensional Yang-Mills 
theory (classical or quantum) in terms of gauge invariant degrees of freedom. 
It is now a theory of metrics on 3-manifolds; which however is not diffeomorphism 
invariant because of the term $g_{ii}$ in the action. As a result, not only the 
geometry of the 3-manifold, but also the metric $g_{ij}$ of any coordinate 
system chosen on the manifold is relevant.

For 3 dimensional (Euclidean) gravity, an integration over $e$ 
(\ref{auxcontpart}) would give the 
$\delta$-function constraint $R_{ij}=0$, resulting in a topological
field theory 
\cite{W}. There are no massless gravitons as a consequence. Now however, the 
diffeomorphisms provide massless degrees of freedom corresponding to gluons. 
They may be described as follows. The 3
manifolds are described by the metric $g_{ij}$ in the coordinate system $x$.
We may choose a new coordinate system $\phi^{A}(x)\;(A=1,2,3)$, 
with a standard form of the metric $G_{AB}[\phi]$. We have
\begin{equation}\label{newmetric}
g_{ij}(x)=\frac{\partial\phi^{A}}{\partial x^{i}}G_{AB}[\phi]
\frac{\partial\phi^{B}}{\partial x^{j}}.
\end{equation}
This gives the form of the action as, 
\begin{equation}\label{metricaction}
S=\int d^{3}x \left[ -\left(\frac{\partial\phi^{A}}{\partial 
x^{i}}G_{AB}[\phi] \frac{\partial\phi^{B}}{\partial x^{j}}\right)
+\frac{i}{2\kappa} \left| \frac{\partial\phi^{A}}{\partial 
x^{i}}\right| \sqrt{G[\phi]} R[\phi]\right],
\end{equation}
where $ \left| \frac{\partial\phi^{A}}{\partial x^{i}}\right| = 
det \left( \frac{\partial\phi^{A}}{\partial x^{i}}\right).$
We identify $\phi^{A}(x)\;(A=1,2,3)$ as the dual gluons. A simple way of 
seeing this is as follows. Note that the second term comes with a factor 
$i=\sqrt{-1}$, whereas the first term does not. In this sense it is analogus 
to the $\theta$-term in QCD which continues to have the factor
$i=\sqrt{-1}$ in the Euclidean version..
Consider a random phase approximation to $Z$. The extrema of the phase factor
correspond to solutions of the the 
vacuum Einstein equations. In this case (3 dimensions), this means that the 
space is flat. Now we may choose the standard form $G_{AB}=\delta_{AB}$.
$\phi^{A}$ now represent arbitrary curvilinear coordinates for that
manifold. Then the 
first term in (\ref{metricaction}) is just $(\nabla \phi^{A})^{2}$. This 
describes three massless scalars. As in section 2 they represent the one
transverse degree of freedom for each color. Thus the gluons are now described 
in terms of gauge invariant, local, scalar degrees of freedom.

In the general case $R\neq 0$, consider normal coordinates 
$\phi^{A}(x)$ at a given point. The metric has the standard form,
\begin{equation}\label{stdmetric}
G_{AB}[\phi]=\delta_{AB} + R_{ABCD}[\phi]\:\phi^{C}\phi^{D}+ \ldots .
\end{equation}
$\phi^{A}$ represents the dual gluons and $R$ the geometric aspects of 
the manifold. Both are degrees of freedom of 2+1 dimensional Yang-Mills
theory.  $\phi^{A}$
are invariant under the Yang-Mills gauge transformations. Thus equation 
 (\ref{metricaction}) describes Yang-Mills dynamics in terms of gauge 
invariant degrees of freedom.
\section{Monopoles}

We now identify the monopoles of Yang-Mills theory in terms of the dual variables.
Monopoles are related to Yang-Mills configurations $\{ A_{i}^{a}(x)\}$ 
with a non-trivial U(1) fibre bundle structure \cite{WuYang}. In such 
configurations, the monopoles are characterized by points with the 
following property \cite{GO}. Consider a surface enclosing a point 
and a set of based loops spanning 
it. Consider eigenvalues of the corresponding Wilson loop operator. 
As one spans the sphere, the eigenvalue changes continuously from zero to 
$2\pi$ instead of coming back to zero. Thus such points 
have topological meaning. Moreover a small change in their position can 
produce a large change in the expectation value of the Wilson loop. 
Therefore we may expect that such points are relevant for confinement, 
even though a semi-classical or dilute gas approximation may not be 
available. Therefore it is important to provide a characterization of 
these monopoles and their interactions with the dual gluons.

In case of 't Hooft-Polyakov monopole, the location of the monopoles is 
given by the zeroes of the Higgs field \cite{AFG}. In pure gauge theory we 
do not have such an explicit Higgs field. 't Hooft \cite{HOO}
has proposed use of a composite Higgs for this case.

We follow a different procedure here. Consider the eigenvalue equation 
of the positive symmetric matrix $B_{i}^{a}(x)B_{i}^{b}(x)=I^{ab}(x)$ for 
each $x$.
\begin{equation}\label{eigenval}
I^{ab}(x)\chi_{a}^{A}(x)=\lambda^{A}(x)\chi_{b}^{A}(x).
\end{equation}
The eigenvalues $\lambda^{A}(x)$, $(A=1,2,3)$ are real and the 
corresponding three eigenfunctions $\chi_{a}^{A}(x)$, $(A=1,2,3)$ form 
an orthonormal set. The monopoles in any Yang-Mills configuration 
$A_{i}^{a}(x)$ can be located in terms of $\chi_{a}^{A}(x)$. We will 
illustrate this explicitly in case of the Prasad-Sommerfield solution 
\cite{PS}. For this $I^{ab}$ has the tensorial form,
\begin{equation}\label{tensor}
I^{ab}(x)=P(r)\delta^{ab} + Q(r)x^{a}x^{b}
\end{equation}
with $P(0)\neq 0$ and finite. At $r=0$, the eigenvalues are triply 
degenerate. Away 
from $r=0$, two eigenvalues are still degenerate, but the third one is 
distinct 
from them. The corresponding eigenfunction (labelled A=1, say) is 
$\chi_{a}^{1}(x)=\hat{x}^{a}$. This precisely has the required behaviour
for the composite Higgs at the center of the monopole \cite{HOO}.

We may regard $\chi_{a}^{A}(x)$ as providing three independent triplets 
of (normalized) Higgs fields. Using them, we may construct three abelian 
gauge fields,
\begin{equation}\label{abelian1}
b_{i}^{A}(x)=\chi_{a}^{A}(x)B_{i}^{a}(x)-\frac{1}{3}\epsilon_{ijk}
\epsilon^{abc}\chi_{a}^{A}D_{j}\chi_{b}^{A}D_{k}\chi_{c}^{A}
\end{equation}
We have 
\begin{equation}\label{abelian2}
b_{i}^{A}(x)=\epsilon_{ijk}\partial_{j}a_{k}^{A}-\frac{1}{3}\epsilon_{ijk}
\epsilon^{abc}\chi_{a}^{A}\partial_{j}\chi_{b}^{A}\partial_{k}\chi_{c}^{A}
\end{equation}
where the three abelian gauge potentials are given by $a_{i}^{A}(x)=
\chi_{a}^{A}(x)A_{i}^{a}(x)$. For each $A=1,2,3$, the second part of the 
right hand side is the topological current for the Poincare-Hopf index 
\cite{AFG}. 
It is the contribution of the magnetic fields due to the monopoles. 
These monopoles are located at points where this index is non-zero.

Since, $\chi_{A}^{a}=\epsilon_{ABC}\epsilon^{abc}\chi_{b}^{B}\chi_{c}^{C}$,
we may rewrite our abelian fields as
\begin{equation}\label{abelian3}
b_{i}^{A}(x)=\epsilon_{ijk}(\partial_{j}a_{k}^{A}(x)+\epsilon^{ABC}c_{j}^{B}
c_{k}^{C})
\end{equation}
where $c_{i}^{A}=\epsilon^{ABC}\chi_{a}^{B}\partial_{i}\chi_{a}^{C}$ has 
the form of a `pure gauge' potential, but is not, because of the 
singularity in $(\chi_{a}^{A})$.

Thus for any configuration $A_{i}^{a}(x)$ of the Yang-Mills potential, 
monopoles may be characterized as the points where the eigenvalues of 
the symmetric matrix $B_{i}^{a}(x)B_{i}^{b}(x)$ become triply degenerate. 
We may use the corresponding eigenfunctions to construct three abelian 
gauge fields with respective monopole sources. Instead of $I^{ab}$, we may
also use the gauge invariant symmetric tensor field
$B_{i}^{a}(x)B_{j}^{a}(x)$ and it's eigenfunctions $\chi_{i}^{A}(x)$. This
provides a gauge invariant description of the monopoles.

We may also use the Ricci tensor $R_{i}^{j}=R_{ik}(x)g^{kj}(x)$ for this
purpose. The three eigenfunctions $\chi_{i}^{A}(x)$, $(A=1,2,3)$ (Ricci
principal directions \cite{E}) provide three orthogonal vector fields for the
3-manifold.  In regions where eigenvalues of $R_{i}^{j}$ are degenerate,
the choice of the vector fields is not unique. One can make any choice
requiring continuity. However {\em{isolated points}} where $R_{i}^{j}$ is
triply degenerate are special, and have topological significance. 
At such points the vector fields are singular. Thus the monopoles 
correspond to the singular points of these vector fields. The index of the 
singular point is the monopole number.

We emphasize that the centers have a topological interpretation
which is independent of the way we construct them.


\section{Interaction of dual gluons with monopoles}

Dual gluons are identified with a coordinate system $\phi^{A}(x)$ 
$A=1,2,3$ on the 3-manifold eqns.(\ref{metricaction})(\ref{stdmetric}). We 
now consider special coordinate systems which are singular at the location 
of the monopole. In case of the Prasad-Sommerfield monopole, the 
correspond to the spherical coordinates ($r,\theta,\phi$) with the 
monopole at the origin. In the general case, we may construct the 
coordinate system as follows. At the site of the monopole, one of the 
eigenfunctions $\chi_{i}^{1}(x)$ say, has the radial behaviour. Then we 
may construct the integral curves of this vector field by solving the 
equations, 
\begin{equation}
\frac{dx^{1}}{\chi_{1}^{1}(x)}=\frac{dx^{2}}{\chi_{2}^{1}(x)}
=\frac{dx^{3}}{\chi_{3}^{1}(x)}. 
\end{equation}
We may choose these integral curves to be the equivalent of the 
$r$-coordinate, i.e. we identify these curves with $\theta=$constant, 
$\phi=$constant curves of the new coordinate system. Consider closed 
surfaces surrounding the monopole which are nowhere tangential to these 
integral curves. A simple choice is just the spherical surfaces. We may 
identify them with the surfaces $r$=constant. (We have not specified the 
$\theta,\phi$ coordinates completely, but this is not required for our 
purpose.)
We thus have a coordinate system $\chi^{A}(x)$ whose coordinate 
singularities correspond to the monopoles. In this coordinate system,
$\int d^{3}x\sqrt{g}R=\int d^{3}x\:(\epsilon_{ijk}\epsilon^{ABC}
\partial_{i}\chi^{A}\partial_{j}\chi^{B}\partial_{k}\chi^{C})
\sqrt{\cal G}(x)R(x)$ where ${\cal G}_{ij}$ is the metric in this 
coordinate system.

Now $\partial_{i}(\epsilon_{ijk}\partial_{j}\chi^{2}\partial_{k}\chi^{3})$ 
is non-zero at $x=x_{0}$ and is related to the monopole charge at 
$x_{0}$ as follows. Let 
$\chi^{A}(x)-\chi^{A}(x_{0})=\rho(x)\hat{\chi}^{A}(x)$ where 
$\hat{\chi}^{A}(x)\hat{\chi}^{A}(x)=1$.
We see that there is a coupling of the field combination $\sqrt{G(x)}R(x)
\rho^{3}(x)$ to the monopole charge density $\partial_{i}k_{i}(x) =m_{i}
\delta^{3}(x_{0})$, where $k_{i}(x)=\epsilon_{ijk}\epsilon^{ABC}
\hat{\chi}^{A}\partial_{j}\hat{\chi}^{B}\partial_{k}\hat{\chi}^{C}$.
Thus a certain combination of the dual gluon $\phi^{A}(x)$ and the 
geometric degree of freedom $R(x)$ couples to the monopoles. In analogy 
to the compact U(1) lattice gauge theory (sec.2), this may be expected 
to give a mass for the dual gluon and hence confinement. There are 
other interactions which are not of topological origin and these are to 
be interpreted as self interactions.

\section{Conclusion}

We have argued that the duality transformation for 2+1 dimensional 
Yang-Mills theory can be carried out in close analogy to the abelian case.
The dual theory has geometric interpretation in terms of 
3-manifolds. We identified the dual gluons with the coordinates of 
the 3-manifolds and monopoles with the coordinate singularities.
We expect that this will provide a new approach for 
understanding quark confinement. 

\section{Acknowledgement}
One of us PM wishes to thank Dr. Elizabeth Gasparim and Dr. Mahan Mitra 
for explanation of several mathematical concepts.

\end{document}